\documentclass[aps,pra,twocolumn,groupedaddress,showpacs]{revtex4}
\usepackage{graphicx}
\usepackage{amsmath,amssymb,amsfonts}
\usepackage{natbib}

\begin{document}

\title{Loss of coherence in double--slit diffraction experiments}

\author{A. S. Sanz}
 \email[Present address: Chemical Physics Theory Group, \\
  Department of Chemistry, University of Toronto, \\
  Toronto, Canada M5S 3H6. \\
  E--mail:]{asanz@chem.utoronto.ca}

\author{F. Borondo}
 \email[E--mail: ]{f.borondo@uam.es}
 \affiliation{Departamento de Qu\'\i mica, C--IX,
  Universidad Aut\'onoma de Madrid,
  Cantoblanco--28049 Madrid, Spain}

\author{M. J. Bastiaans}
 \email[E--mail: ]{m.j.baastians@tue.nl}
 \affiliation{Technische Universiteit Eindhoven,
  Faculteit Elektrotechniek, \\
  Postbus 513, 5600 MB Eindhoven, The Netherlands}

\date{\today}

\begin{abstract}
The effects of incoherence and decoherence in the double--slit
experiment are studied using both optical and
quantum--phenomenological models. The results are compared with
experimental data obtained with cold neutrons.
\end{abstract}


\maketitle


\section{\label{sec1} Introduction}

Coherence is an important building block in physics, common to
optics and quantum mechanics.
In classical optics, it simply implies a phase condition related to
interference.
Its lack is related to the nature of the source of light
(non--monocromaticity) and/or the geometrical shape of the objects
placed on the source--detector pathway \cite{Wolf1,Wolf2}.
In this sense, incoherence can be regarded as a ``non dynamical''
effect.
On the other hand, the same concept acquires a more general meaning
in quantum mechanics, where it has to reconcile with the statistical
interpretation of the wave function.
When two or more states are coherently superposed their properties
qualitatively differ from those exhibited by the isolated components,
this being a distinctive feature of the quantum behavior
\cite{Ballentine1}.

Due to its remarkable relevance in the modern quantum theory of
information and quantum computation \cite{Nielsen}, a topic of
much recent interest is the understanding of how quantum systems
lose their coherence, being decoherence the most widely accepted
mechanism \cite{Giulini1}.
By decoherence we understand the irreversible emergence of classical
properties in a quantum system through its ``dynamical'' interaction
with an environment or bath \cite{Kiefer1,Kiefer2}.
This idea is better understood by splitting the Hamiltonian describing
the full system in three terms:
\begin{equation}
  \hat{H} = \hat{H}_A + \hat{H}_B + \hat{V}_{AB} ,
 \label{eq:1}
\end{equation}
where $\hat{H}_A$ and $\hat{H}_B$ are the Hamiltonians governing
the evolution of system and environment, respectively, and
$\hat{V}_{AB}$ accounts for the coupling between them.
This last term leads to a fast cancellation of the off--diagonal
elements of the system reduced density matrix, which is studied
by averaging (i.e., tracing) the full--system density matrix over
the environment degrees of freedom.

In general, the visibility of an experimental diffraction pattern
can be considered the result of combining incoherence and decoherence.
In this paper we analyze the influence of these two mechanisms using
a theoretical model applied to the double--slit experiment with cold
neutrons carried out by Zeilinger {\it et al.} \cite{Zeilinger1}.
By means of a detailed optical study of the neutron beam \cite{Wolf1},
we conclude that incoherence and decoherence are both needed in order
to explain the loss of coherence found in the experiment, this being
the main result of our work.
Decoherence is introduced here using a simple phenomenological theoretical
model that assumes an exponential damping of the interferences
\cite{Savage}, and that only has one empirical parameter (the coherence
degree).
It should be stressed out that this model is based solely on first
principles (except for the estimation of the coherence parameter).
Other authors have treated the effects of incoherence in similar diffraction
experiments with He atoms \cite{Toennies1} and He clusters \cite{Toennies2}
from an optical point of view, and also the effects of decoherence in
fullerene diffraction \cite{Viale}.

The organization of the paper is as follows.
To make the paper self--contained, we briefly review in Sec.~\ref{sec2}
the experiment carried out by Zeilinger {\it et al.}\ \cite{Zeilinger1}.
Section \ref{sec3} is devoted to the optical description of this experiment,
and in Sec.~\ref{sec4} we present our quantum calculations,
whose results are compared with the experimental data.
Finally, the main conclusions derived from this work are summarized in
Sec.~\ref{sec5}.


\section{\label{sec2} The experiment}

The double--slit used by Zeilinger {\it et al.}~\cite{Zeilinger1}
consisted of a highly absorbing boron wire mounted into the
opening gap of a single--slit.
The dimensions of this arrangement (hereafter labelled $O$) are
$a_1$--$d$--$a_2 =$ 21.9--104.1--22.5~$\mu$m
(left slit/boron wire/right slit).
The wavelength of the neutron beam, $\lambda_{\rm dB} = 18.45$~\AA\
(with a bandwidth $\Delta \lambda = \pm 1.40$~\AA), was selected
by means of an entrance single--slit ($C$) of width $w = 20$~$\mu$m,
located at a distance $z = 5$~m before $O$.
After crossing the double--slit and a pathway $v = 5$~m,
neutrons are made to pass through a scanning single--slit ($D$)
of width $w_0 = 20$~$\mu$m before reaching the detector
(located behind $D$).

\begin{figure}
\includegraphics[width=8.5cm]{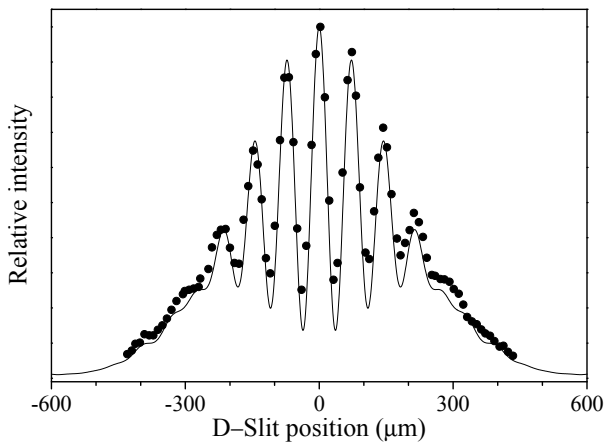}
 \caption{\label{fig:1}
  Experimental results obtained by Zeilinger \emph{et al.}
  \protect\cite{Zeilinger1} for the double--slit diffraction
  of cold neutrons (solid circles), and
  theoretical intensity patterns obtained by assuming
  two finite--size, identical slits [see Eq.~(\ref{eq:15})]
  (solid line).}
\end{figure}

The reported results are reproduced in Fig.~\ref{fig:1}.
From them, a value for the fringe visibility of
$\mathcal{V}_{\rm exp}=0.583$ is obtained.
This magnitude is defined \cite{Wolf1,Wolf2} as:
\begin{equation}
 \mathcal{V} =
  \frac{I_{\rm max} - I_{\rm min}}{I_{\rm max} + I_{\rm min}} ,
 \label{eq:2}
\end{equation}
being $I_{\rm max}$ and $I_{\rm min}$ the intensities corresponding
to the central maximum and the first minimum beside it,
respectively.
This experimental lack of visibility cannot be attributed to simple
causes, such as non--convergence of the intensity pattern at $D$
or the size difference between slits.
In the first case, $D$ is far enough from $O$, and the intensity
pattern is well--defined (it displays almost vanishing minima).
In the second case, the difference between $a_1$ and $a_2$ is too
small ($\simeq 3\%$), and the incoherence effects due to this asymmetry
are irrelevant.
In the following sections the physical factors leading to this high
value for the loss of coherence in the experiment are examined.

\section{\label{sec3} The optical treatment}

Studies on neutron interferometry can be accurately carried out by
means of classical optics \cite{Rauch} provided that energy is low
and spin effects are negligible.
Since classical optics only accounts for incoherence effects,
this approach is of great help in our case in order to discern
whether or not decoherence effects are relevant.
This analysis will also provide the physical conditions that the
quantum model for the possible decoherence must fulfill.
Therefore, we first analyze here the optics of the neutron beam
to establish the relevance of incoherence effects in the diffraction
pattern.

The coherence of light can be understood as the effect of stationary
stochastic processes \cite{note1}, i.e., random fluctuations in
the amplitude and phase of the interfering waves of light in time.
Although the source of that stochasticity is irrelevant for the optical
analysis, it turns out to be crucial in the quantum theory of decoherence
\cite{Giulini1}.
The correlation between two processes taking place at different points and
times is given by the (mutual) coherence function \cite{Wolf3,Pap68,Bas77},
$\Gamma(x_1,x_2;\tau)$, which is only a function of the time difference,
$\tau = t_1-t_2$, due to stationarity.
From it, the power spectrum \cite{Pap68,Bas77} or cross--spectral density
function \cite{Man76}
\begin{equation}
 S(x_1,x_2;\omega) = \int \Gamma(x_1,x_2;\tau)
   {\rm e}^{-{\rm i}\omega\tau} d\tau
 \label{eq:3}
\end{equation}
can be defined.
This is the optical analog of the quantum density matrix.
Like it, $S(x_1,x_2;\omega) = S^*(x_2,x_1;\omega)$ for the off--diagonal
elements, while the diagonal ones, $S(x,x;\omega)$, are real and
give the intensity distribution of light at a frequency $\omega$.
Therefore, the power spectrum is the essential element in this
optical treatment.

Since Eq.~(\ref{eq:3}) refers to the neutron beam, $\omega$ can be
substituted by $\lambda$ without loss of generality.
As in Ref.~\cite{Zeilinger1}, in the theoretical description of the
experiment we will take into account the pass of the neutron beam through
$C$, $O$, and $D$.
If $C$ is identified with a neutron source that can be assumed as
incoherent \cite{Zeilinger1}, i.e., the intensity at each point along
the aperture does not depend on any other point, then
\begin{equation}
 S_C(x_1,x_2;\lambda) = q(x_1;\lambda) \delta(x_1 - x_2) .
 \label{eq:4}
\end{equation}
Here, $q(x,\lambda) \geq 0$ is the average intensity at $x$.
When no absorption takes place at the borders of $C$
(i.e., each point of the aperture produces the same intensity)
and $\lambda$ is independent on the position along this aperture,
$q(x;\lambda)$ is given by
\begin{equation}
 q(x;\lambda) = \frac{1}{w} s(\lambda) , \qquad |x| \le w/2,
 \label{eq:5}
\end{equation}
with $s(\lambda)$ describing the spectral profile of the neutron beam.
It should be emphasized that to keep the explicit dependence on $\lambda$
is important, since the source is not totally monochromatic.
As will be seen below, the spectral profile plays a key role on the shape
of the measured intensity at $D$ within this optical treatment.

From $C$ to $O$ the neutron beam propagates freely according to
the following point--spread function \cite{Pap68}:
\begin{equation}
 h_{CO}(x,\xi) = {\rm e}^{{\rm i}k(x-\xi)^2/2z} ,
 \label{eq:6}
\end{equation}
where $x$ and $\xi$ refer to the spatial coordinates at $O$ and $C$
\cite{note2}, respectively, and $k = 2\pi/\lambda$.
In this way, the power spectrum just before $O$ becomes
\begin{multline}
 S_O'(x_1,x_2;\lambda) = \\
   \int h_{CO}(x_1,\xi_1) h_{CO}^*(x_2,\xi_2)
        S_C(\xi_1,\xi_2;\lambda) d\xi_1 d\xi_2 \\
   = \bar{q}(\Delta x;\lambda) \
        {\rm e}^{{\rm i}k(x_1^2-x_2^2)/2z} ,
 \label{eq:7}
\end{multline}
where
\begin{equation}
 \bar{q}(\Delta x;\lambda) = {\rm sinc}[k(x_1-x_2)w/2z] s(\lambda) ,
 \label{eq:8}
\end{equation}
with $\Delta x = x_1 - x_2$ and ${\rm sinc}\beta = \sin\beta/\beta$.
As can be seen, the finite size of the source gives rise to a coherence
in the neutron beam, which manifests as a typical single--slit
diffraction pattern at $O$.
The first minima of this correlation pattern \cite{note3} take place
at $\Delta x = \pm z\lambda_{\rm dB}/w \simeq \pm 461$~$\mu$m.
Therefore, the spatial width illuminated by its central maximum will
be relatively larger than the extension covered by the double--slit
arrangement.
This means that the two diffracted beams will display a slight divergence,
a fact that must be included as a source of incoherence in the quantum
model (see Sec.~\ref{sec4}).

The double--slit acts as a modulator of the incoming wave with a
modulation function $m(x)$. Thus, the power spectrum at $O$ can be
written as
\begin{equation}
 S_O (x_1,x_2;\lambda) = m(x_1) m^*(x_2) S_O' (x_1,x_2;\lambda) .
 \label{eq:9}
\end{equation}
After crossing the double--slit the two diffracted beams described
by $S_O$ propagate freely until reaching $D$.
There, the power spectrum, $S_D(x_1,x_2;\lambda)$, is computed by
replacing, in Eq.~(\ref{eq:7}), $h_{CO}(x,\xi)$ and
$S_C(\xi_1,\xi_2;\lambda)$ by $h_{OD}(x,\xi)$ and
$S_O(\xi_1,\xi_2;\lambda)$, respectively.
The diagonal elements of $S_D$ give the intensity distribution,
$I(x;\lambda)$, right before the neutron beam crosses $D$.
For example, in the case of two identical very narrow slits
(such that the modulation function can be expressed as the sum of two
$\delta$--functions) separated from each other a distance
$\bar{d} = d + (a_1 + a_2)/2$, one obtains
\begin{multline}
 S_D(x_1,x_2;\lambda) =
  \Big[ \cos (k \Delta \bar{d}/2v) s(\lambda) \\
 + \bar{q}(\bar{d};\lambda) \cos [k (x_1 + x_2) \bar{d}/2v] \Big]
  {\rm e}^{{\rm i}k(x_1^2-x_2^2)/2v} ,
 \label{eq:10}
\end{multline}
from which
\begin{equation}
 I(x;\lambda) = \Big[ 1 +
  {\rm sinc}(k\bar{d}w/2z) \cos (k\bar{d}x/v) \Big] s(\lambda) .
 \label{eq:11}
\end{equation}
The separation between two consecutive maxima that results from
(\ref{eq:11}) is $v\lambda_{\rm dB}/\bar{d} \simeq 73$ $\mu$m, which
is in good agreement with the experimental value of Fig.~\ref{fig:1}.
However, the value of the fringe visibility, $\mathcal{V} = 0.881$, is
still far from the experimental value.
Moreover, the amplitude of the oscillations is constant, in strong
disagreement with the experimental pattern.

If the finite size of $D$ is taken into account, the measured
intensity for a position $x$ of the detector results
\begin{equation}
 I_0(x;\lambda) = \frac{1}{w_0}
  \int_{x - w_0/2}^{x + w_0/2} I(x';\lambda) dx' ,
 \label{eq:12}
\end{equation}
which decreases the fringe visibility, but leaves the profile of
the interference pattern unaffected.
The non--monochromaticity of the neutron beam leads to a subsequent
integration over the spectral bandwidth.
Taking into account the optical relation $k = \omega/c$, this integral
can be carried out easily by assuming $\omega = 2\pi c/\lambda$,
and substituting the corresponding quantities by $\lambda_{\rm dB}$
and $\Delta \lambda = 2.80$~\AA.
Thus, in the particular case of a uniform spectral profile over the
bandwidth $\Delta \omega$, one obtains
\begin{multline}
 I_0(x) = \frac{1}{\Delta \omega}
  \int_{\omega - \Delta \omega/2}^{\omega + \Delta \omega/2}
    I_0(x;\omega') d\omega' \\
  \propto 1 + {\rm sinc}(k\bar{d}w/2 z)
  \ {\rm sinc} (k\bar{d}w_0/2 v) \\
  \times
   {\rm sinc}[(\Delta\lambda/\lambda_{\rm dB})(k\bar{d}x/2v)]
    \cos(k\bar{d}x/v) ,
 \label{eq:13}
\end{multline}
where the approximation of constant sinc--functions (with $\lambda
= \lambda_{\rm dB}$) has been used, since these functions vary
slowly within the integration interval (about a 4\% with respect
to their value at $\lambda = \lambda_{\rm dB}$).
The results obtained with this expression are much better than
those for Eq.~(\ref{eq:11}) but still far from the experimental
data.
In particular a value of $\mathcal{V} = 0.772$ is obtained,
and the averaging process introduces a certain modulation
with respect to the constant--amplitude intensity of Eq.~(\ref{eq:11}).
However, the decay in the intensity pattern as the absolute value of
the detector position increases is not satisfactory.
This is a general feature [i.e., independent of the spectral profile
determined by $s(\lambda)$] coming from the assumption of point slits
at $O$.

The modulation of the intensity pattern is a border effect caused
by the finite size of the slits, while the oscillations are the
result of the interference of two spatially separated wavefronts.
This can be easily shown by considering that the slits are
coherently illuminated (i.e., the modulation function is described
by the sum of two hat functions), and their size is relatively small
\cite{note4}.
In this way, one can assume that Eq.~(\ref{eq:11}) still holds,
and introduces the sinc--function caused by a finite--size slit into it,
thus obtaining
\begin{multline}
 I(x,\lambda)
 \propto\ \Big[ {\rm sinc}^2 (k\bar{a}\eta^-/2v)
  + {\rm sinc}^2 (k\bar{a}\eta^+/2v) \\
  + \ 2 \ {\rm sinc}(k\bar{a}\eta^-/2v) \ {\rm sinc}(k\bar{a}\eta^-/2v) \\
  \times {\rm sinc}(k\bar{d}w/2z) \cos (k\bar{d}x/v) \Big]
    s(\lambda) .
 \label{eq:14}
\end{multline}
Here, for the sake of simplicity, both slits are considered with the same
width, $\bar{a} = (a_1 + a_2)/2$, and $\eta^\pm = x \pm v\bar{d}/2b$.
Using the same arguments, an equivalent expression for Eq.~(\ref{eq:13})
can be calculated,
\begin{multline}
 I_0(x) \propto \
  {\rm sinc}^2 (k\bar{a}\eta^-/2v) + {\rm sinc}^2 (k\bar{a}\eta^+/2v) \\
  + \ 2 \ {\rm sinc}(k\bar{a}\eta^-/2v) \ {\rm sinc}(k\bar{a}\eta^-/2v)
   \ {\rm sinc}(k\bar{d}w/2z) \\
   \times {\rm sinc}(k\bar{d}w_0/2v) \
   {\rm sinc}[(\Delta\lambda/\lambda_{\rm dB})(k\bar{d}x/2v)]
  \cos (k\bar{d}x/v) ,
 \label{eq:15}
\end{multline}
The corresponding results are shown as solid line in Fig.~\ref{fig:1}.
As can be seen, the intensity pattern fits much better the experimental
results, presenting only a small discrepancy for the central minima.
However, the fringe visibility, $\mathcal{V} = 0.760$, does not still
fully account for the value obtained experimentally, although it
improves with respect to the previous cases.
Therefore, the most important conclusion derived from this phenomenological
analysis of the experiment is that, although different causes contribute to
the incoherence of the neutron beam, they are not sufficient to explain the
experimental results.
The remaining contribution should then be attributed to dynamical causes,
i.e., decoherence, which will be analyzed in the next section.

One final consideration is worth mentioning.
The finite size of the slits is an important factor to take into account
in the formulation of the quantum model.
Indeed, a general phenomenological form for the measured intensity
accounting for the loss of coherence can be postulated from
Eq.~(\ref{eq:15}) as
\begin{equation}
 \mathcal{I}(x) =
  \mathcal{I}_1(x) + \mathcal{I}_2(x)
   + 2 \mathcal{A}\ \! \mathcal{I}_{12}(x) \cos (k\bar{d}x/v) ,
 \label{eq:16}
\end{equation}
where $\mathcal{I}_i(x)$ is the intensity function produced by the
slit $i$ (with $i = 1,2$), $\mathcal{I}_{12}(x)$ the intensity
function modulating the interference, $\mathcal{A}$ the coherence
degree or damping coefficient, and the cos--function is the term
purely due to the interference between the two slits.
Notice that coherence degree and fringe visibility are not the same
unless both partial waves contribute equally to the intensity,
as happens in Eq.~(\ref{eq:11}), where
$\mathcal{I}_1(x) = \mathcal{I}_2(x) = \mathcal{I}_{12}(x) = 1/2$.


\section{\label{sec4} The quantum treatment}

Optical models are attractive because of their their simplicity
(and accuracy when used in association with fitting procedures).
However, they do not deal with the loss of coherence due to
dynamical processes, such as those taking place in the regions
where the neutron beam evolves freely.
The effect produced by such processes is what we call decoherence.
In general, decoherence is understood as a complex effect concerning
the phase of a many--body wave function \cite{Omnes}.
For example, according to the model proposed by Caldeira and Leggett
\cite{Caldeira1}, decoherence arises from the interaction of the system
with a bath of harmonic oscillators linearly coupled to the system.
The collisions between particles belonging to the system and the
environment constitute another mechanism proposed to explain the
loss of coherence \cite{Kiefer1,Kiefer2} in physical processes.
Although these scattering events may be individually quite inefficient,
when taking place in large numbers they may lead to an exponential damping
of the off--diagonal elements of the system reduced density matrix, since
\begin{equation}
 \frac{\partial \tilde{\rho}_{nm}}{\partial t}
   \bigg\arrowvert_{\rm scattering}
   = - \lambda \tilde{\rho}_{nm} (t) ,
 \label{eq:17}
\end{equation}
where $\lambda=\Gamma(1-\langle\psi_0|S_m^\dagger S_n|\psi_0\rangle )$,
$\Gamma$ is the collision rate, and the scattering process between states
$|n\rangle$ and $|m\rangle$ is described by the corresponding $S$--matrix
element.

According to the physical considerations made in the previous section,
in our quantum--mechanical description of the process we only consider
the evolution of the neutron beam from $O$ to $D$, assuming that both
outgoing partial waves at $O$ are coherent (i.e., finite--size slits).
Moreover, as mentioned above, these partial waves differ in width
(because of the different slit sizes), and then move apart separating
along the two opposite directions (notice, however, that the component
of this motion is relatively small).
These initial conditions are the responsible for incoherence.
Taking this into account, the evolution of the neutron beam after passing
through $O$ (and without being affected by an external environment)
can be described at any subsequent time \cite{note5} as
\begin{equation}
 |\Psi^{(0)}\rangle_t =
  c_1 |\psi_1\rangle_t + c_2 |\psi_2\rangle_t ,
 \label{eq:18}
\end{equation}
where $|\psi_i\rangle_t$ is the partial wave emerging from slit $i$,
and $|c_1|^2 + |c_2|^2 = 1$ at any time.
In the coordinates representation, the density matrix associated to this
wave function is
\begin{equation}
 \rho_t^{(0)} ({\bf r}, {\bf r}') =
  \Psi^{(0)}_t ({\bf r}) \left[ \Psi^{(0)}_t ({\bf r}') \right]^* ,
 \label{eq:19}
\end{equation}
with $\Psi^{(0)}_t({\bf r})=\langle {\bf r}|\Psi^{(0)}\rangle_t$.
The diagonal terms of (\ref{eq:19}) give the probability density
or measured intensity
\begin{multline}
 \rho_t^{(0)} = \rho_t^{(0)} ({\bf r}, {\bf r}) \\ =
  |c_1|^2 |\psi_1|_t^2 + |c_2|^2 |\psi_2|_t^2
   + 2 |c_1| |c_2| |\psi_1|_t |\psi_2|_t \cos \delta_t ,
 \label{eq:20}
\end{multline}
where $\delta_t$ is the time--dependent phase difference between
both partial waves.

When the action of an environment is considered, the wave function
(\ref{eq:18}) does no longer describe the evolution of the full system.
Since our main interest focuses on the question of how the environment
affects the coherence of the system by gradually suppressing the
(non--classical) oscillatory term in Eq.~(\ref{eq:20}),
we will assume conditions of elastic scattering \cite{Breuer}.
Accordingly, only the states describing the environment change during
the scattering process, while the system states remain unchanged.
In particular, in our case both partial waves will propagate as free
Gaussian wave packets since they are Gaussians initially (see below).
In this way, the initial coherent state
\begin{equation}
 |\Psi \rangle = |\Psi^{(0)}\rangle \otimes |E_0\rangle
 \label{eq:21}
\end{equation}
becomes an entangled state with the form
\begin{equation}
 |\Psi \rangle_t =
    c_1 |\psi_1\rangle_t \otimes |E_1\rangle_t
  + c_2 |\psi_2\rangle_t \otimes |E_2\rangle_t
 \label{eq:22}
\end{equation}
at any subsequent time.
Here, $|\psi_1\rangle$ and $|\psi_2\rangle$ are coherent wave packets
(or superposition of them),
\begin{multline}
 G (x,z) \propto
  {\rm e}^{- (x - x_0)^2/4 \sigma_x^2 + {\rm i} p_x x} \\
   \times {\rm e}^{- (z - z_0)^2/4 \sigma_z^2 + {\rm i} p_z z} ,
 \label{eq:23}
\end{multline}
where $x$ ($z$) is the parallel (perpendicular) coordinate
(with respect to the plane of the double--slit),
$x_0$ ($z_0$) the center of the wave packet along the $x$ ($z$) axis,
$p_x$ ($p_z$) the parallel (perpendicular) component of the momentum,
and $\sigma_x$ ($\sigma_z$) the width of the wave packet along $x$ ($z$).
In the case of quasi--plane waves, the corresponding wave function can be
constructed as a coherent superposition of Gaussian wave packets distributed
over the width covered by each slit.
On the other hand, notice from Eq.~(\ref{eq:22}) that, due to the initial
coherence, $|E_1\rangle = |E_2\rangle = |E_0\rangle$.
In our case, this initial state for the environment is chosen as
$|E_0\rangle = |\mathbb{I}\rangle$.

In order to calculate the diffraction intensity, the reduced density
matrix for the system is obtained by tracing the full--system density
matrix over the environment states,
\begin{equation}
 \hat{\tilde{\rho}}_t =
  \sum_i \ _t\langle E_i | \hat{\rho}_t | E_i \rangle_t .
 \label{eq:24}
\end{equation}
In the coordinate representation this operation is equivalent to carry out
the integral of the total density matrix over all the 3$N$ degrees of freedom,
$\{ {\bf r}_i \}_{i = 1}^N$, of the environment,
\begin{multline}
 \tilde{\rho}_t ({\bf r}, {\bf r}') = \\
  \int \langle {\bf r}, {\bf r}_1, \ldots {\bf r}_n |
  \Psi \rangle_t \ _t\langle \Psi | {\bf r}', {\bf r}_1,
  \ldots {\bf r}_n \rangle \ \! {\rm d}{\bf r}_1
    \cdots {\rm d}{\bf r}_n .
 \label{eq:25}
\end{multline}
By substituting Eq.~(\ref{eq:22}) into Eq.~(\ref{eq:24}), one obtains then
the reduced density matrix:
\begin{multline}
 \tilde{\rho}_t ({\bf r}, {\bf r}') = \\
 (1 + |\alpha_t|^2) \Big[ |c_1|^2 \psi_{1t} ({\bf r}) \psi_{1t}^* ({\bf r}')
  + |c_2|^2 \psi_{2t} ({\bf r}) \psi_{2t}^* ({\bf r}') \Big]
  \\ +
  2 \alpha_t c_1 c_2^* \psi_{1t} ({\bf r}) \psi_{2t}^* ({\bf r}')
  + 2 \alpha_t^* c_2 c_1^* \psi_{2t} ({\bf r}) \psi_{1t}^* ({\bf r}') ,
 \label{eq:26}
\end{multline}
where $\alpha_t = \!\!\! \ _t\langle E_2 | E_1 \rangle_t$.
From Eq.~(\ref{eq:26}) we obtain the measured intensity
\begin{multline}
 \tilde{\rho}_t = (1 + |\alpha_t|^2)
  \\ \times \Big[ |c_1|^2 |\psi_1|_t^2 + |c_2|^2 |\psi_2|_t^2 +
  2 \Lambda_t |c_1| |c_2| |\psi_1|_t |\psi_2|_t \cos \delta'_t \Big] ,
 \label{eq:27}
\end{multline}
with
\begin{equation}
 \Lambda_t = \frac{2 |\alpha_t|}{(1 + |\alpha_t|^2)} .
 \label{eq:28}
\end{equation}
For the sake of simplicity, we have assumed that the phase difference
between the environment states (included in $\delta'_t$) is constant.
Equation (\ref{eq:27}) is the quantum analog of the optical
(phenomenological) equation~(\ref{eq:16}).
Thus, $\Lambda_t$ is defined as the quantum coherence degree,
which plays here the same role as $\mathcal{A}$ in Eq.~(\ref{eq:16}).

Now, we make the (reasonable) assumption that the environment states
are too complex to keep mutual coherence as time increases \cite{Omnes2}.
In this way, even if they are initially coherent, they become orthogonal
as time passes.
Thus, one can assume $|\alpha_t| \simeq {\rm e}^{-t/\tau_c}$,
being $\tau_c$ the coherence time, a measure of how fast the system
becomes classical (smaller $\tau_c$ implies faster loss of coherence).
By substituting the value of $|\alpha_t|$ in Eq.~(\ref{eq:28}) one obtains
\begin{equation}
 \Lambda_t = {\rm sech} (t/\tau_c) ,
 \label{eq:29}
\end{equation}
which establishes a relationship between coherence degree and
coherence time.
Although the value of the coherence time can be derived analytically
for interfering waves by means of simple Markovian models \cite{Savage},
Eq.~(\ref{eq:29}) allows to determine it from the coherence degree.
In our case, we have used $\Lambda_t = 0.63$ (or, equivalently,
a coherence time of $\tau_c \simeq 5.08 \! \times \! 10^{-2}$~s),
for which an excellent agreement between the experimental data and
the theoretical results is found (see below).
This value corresponds to a fringe visibility $\mathcal{V} = 0.607$
(notice that $\Lambda_t \simeq \mathcal{V}$, as said at the end of
Sec.~\ref{sec3}), very close to the experimental value
($\sim 4\%$ above this value).

Taking all these ingredients into account, we present next results for two
basic models.
In the first one, after passing through the corresponding slits,
$|\Psi_1|^2$ and $|\Psi_2|^2$ are considered to be quasi--plane waves;
while in the second one, their profiles are modelled with Gaussian
functions.
We also assume that the slits are on the $XY$--plane, the direction
of propagation of the beam is along $Z$, and $\ell_y\gg\ell_x$
($\ell_x$ and $\ell_y$ being the dimensions of the slits).
In this way, the model can be considered as two--dimensional,
with the motion taking place in the $X\!Z$--plane.
In order to minimize the effects of the spreading of the wave function
along $z$ with time, given by the general expression
\begin{equation}
 \sigma_t = \sigma_0 \
  \sqrt{ 1 + \left( \frac{\hbar t}{m \sigma_0^2} \right)^2 } ,
 \label{eq:30}
\end{equation}
it is enough, in our case, to choose $\sigma_0^z=2\bar{a}$, what
ensures $\sigma_t \simeq \sigma_0$ during the time of propagation.


\subsection{Case I: Quasi--plane waves}

In this first case, the probability density at the left slit has been
simulated by covering its width with 30 Gaussian packets,
while for the right (wider) one 31 packets have been used.
For both slits $z_0=0$ for all Gaussians, and the distance between
neighbor packets has been taken equal to a constant,
$\Delta_i = a_i/(N_i - 1)$, with $i = 1, 2$.
The width parameter along the $X$--direction is $\sigma_x^{(i)}=a_i/N_i$,
with $N_i = 30$ or 31 depending on the slit considered,
while for the $Z$--direction $\sigma_z = 2 \bar{a}$, as explained above.
The time evolution of the wave function has been numerically obtained
using Heller's method \cite{Sanz2}, which is exact in our case.

\begin{figure}
 \includegraphics[width=8cm]{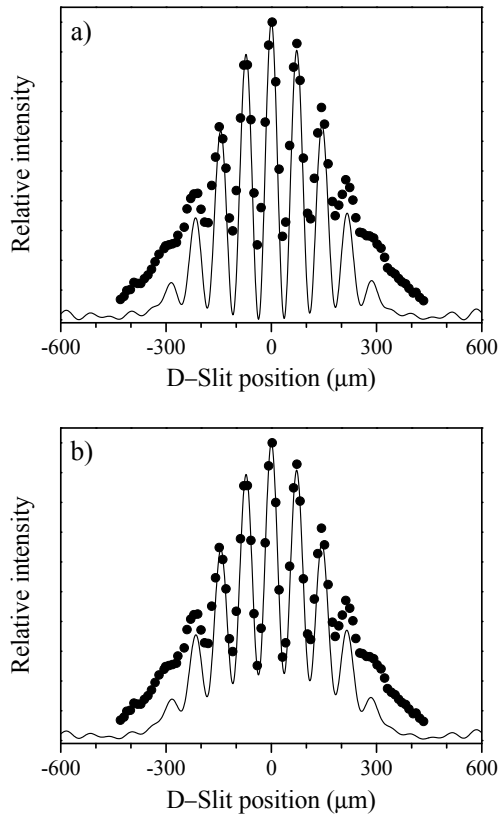}
 \caption{\label{fig:2}
  Intensity obtained from the interference of two quasi--plane waves
  in the case of:
  (a) only incoherence effects, and
  (b) incoherence plus decoherence effects.
  For comparison,   the experimental data (solid circle) are also shown.}
\end{figure}

In Fig.~\ref{fig:2}(a) we show the results corresponding to a case
where only incoherence effects are included [Eq.~\ref{eq:20}].
As mention above, after passing through the slits each partial wave has
a certain motion component along the perpendicular direction.
The (initial) momentum component along each direction can be estimated
using the uncertainty principle, according to which
$p_x \sim \Delta p_x \sim \hbar/\Delta x$.
Since $\Delta x \sim a$, the momentum along the parallel direction
will be $p_x^{(i)} \simeq \pm \hbar/a_i$, with the plus (minus) sign
corresponding to the right (left) slit.
Hence, the momentum along the $Z$--direction is
$p_z^{(i)}=[(2\pi\hbar/\lambda_{\rm dB})^2-(p_x^{(i)})^2]^{1/2}$.
As can be seen in the figure, the results indicate that the effects due to
incoherence are very small, with the minima of the oscillations almost
touching the horizontal axis; they display, however, a certain degree of
asymmetry with respect to $x=0$.
The problem results quite similar to the evolution of the same initial wave
with both slits of the same width and no perpendicular motion.
Finally, notice the oscillations that appear for $x \gtrsim 400$ \AA.
They arise from the diffractive effects caused by the borders of the slits,
and are not observed in the equivalent solid curve of Fig.~\ref{fig:1}
because of the double averaging process (with respect to the finite--size
of $D$ and the bandwidth) carried out there.
As will be seen in next section, these oscillations disappear when
a model based on Gaussians is used.

To conclude this section, we show in Fig.~\ref{fig:2}(b) the intensity
resulting when both incoherence and decoherence (with $\Lambda_t = 0.63$)
effects are included [Eq.~(\ref{eq:27})].
As can be seen, these results have been greatly improved with respect to
only consider incoherence, showing an important decrease in the fringe
visibility.
However, the agreement with the experimental data is still poor.
It should be stressed out that they have been obtained by means of a
reasonable theory based on first principles, with no averaging as
in Sec.~\ref{sec3}.


\subsection{Case II: Gaussian waves}

In this section we show that the results presented in the previous
section can be highly improved by simply considering that the
slits are Gaussian \cite{Feynman1}, i.e., the wave function in
Eq.~(\ref{eq:18}) is a linear superposition of two Gaussian wave
packets.
These Gaussian wave packets are centered at $x_0^{(1)}=(a_1-d)/2$
and $x_0^{(2)} = (a_2 + d)/2$ (with $z_0^{(i)} = 0$),
respectively, and their corresponding width parameters are taken as
$\sigma_x^{(i)} = a_i/4$ and $\sigma_z = 2a$.
The value of $\sigma_x^{(i)}$ implies that the intensity at the border
of the corresponding slit is
$|\psi_i (\pm \sigma_x^{(i)}/2)|^2/|\psi_i (x_i)|^2 = {\rm e}^{-2}$
when $(x - x_i) = \sigma_x^{(i)}/2$, i.e., approximately 13.5\%
of the total intensity.
Thus, only a very small part of the wave will be out of the boundaries
defined by the borders of the slits.
This assumption is in agreement with the experimental observations,
since Zeilinger {\it et al.} reported, with regard to the error on
the slit widths, that ``neutrons penetrating through the boron
wire along a chord 0.2 $\mu$m away from the surface are attenuated
by more than a factor $1/e$'', such that these ``neutrons would be
refracted far out of the diffraction pattern'' \cite{Zeilinger1}.

\begin{figure}
 \includegraphics[width=8cm]{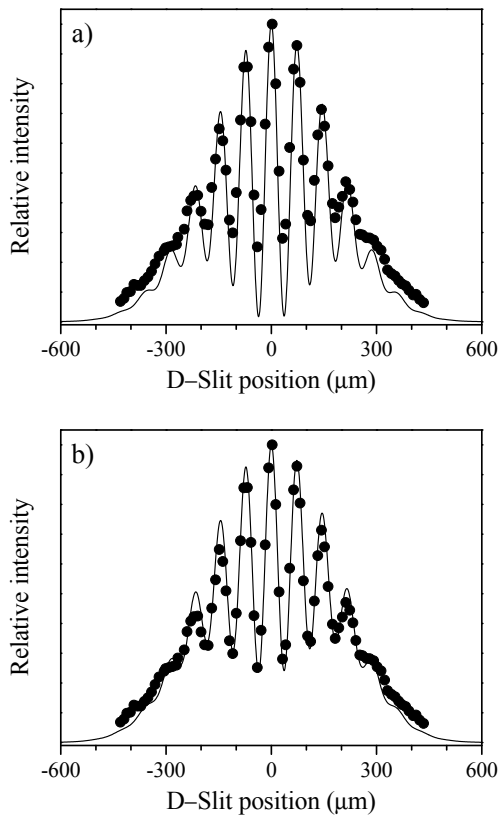}
 \caption{\label{fig:3}
  Intensity obtained from the interference of two
  Gaussian slits for: (a) only incoherence effects, and
  (b) incoherence plus decoherence effects. For comparison,
  the experimental data (solid circle) are also shown.}
\end{figure}

In sharp contrast to the case described in the previous subsection,
when incoherence is considered in this model (given, as above,
by $p_x^{(i)} \simeq \pm \hbar/a_i$ and $p_z^{(i)} =
[(2\pi\hbar/\lambda_{\rm dB})^2 - (p_x^{(i)})^2]^{1/2}$,
for slits with different sizes), the resulting intensity,
shown in Fig.~\ref{fig:3}(a), looks quite different compared with
the results of Fig.~\ref{fig:2}(a).
In this case, it can be observed how the Gaussian envelope due to
single--slit diffraction (which modulates the amplitude of the
intensity maxima) allows the interference pattern to fit the shape
of the experimental data fairly well.
Nonetheless, the central minima do not completely agree with the
experimental ones yet.
This theoretical pattern looks very much like that given by
Eq.~(\ref{eq:15}) (see results in Fig.~\ref{fig:1}).
Notice that the diffractive effects have disappeared, and then
the ``tail'' of the curve decreases exponentially, as expected.
Qualitatively, we can state that this model reproduces the experimental
pattern is as good as the optical theory, but with the advantage that
here we do not need to assume any average.

Finally, when we consider Eq.~(\ref{eq:27}) with $\Lambda_t = 0.63$,
which includes both incoherence and decoherence effects in this
Gaussian--slit approach, the agreement between theoretical and
experimental results is excellent, as can be seen in Fig.~\ref{fig:3}(b).
Therefore, this experiment cannot only be interpreted in terms of
optical models, but needs to include decoherence as a mechanism
leading to the loss of coherence.
This is not an \emph{a priori} expected result since the neutron mass
is very small.


\section{\label{sec5} Conclusions}

Optical models are currently used in slits diffraction experiments
with particles like electrons, neutrons, atoms or fullerenes
in order to explain the behavior of the empirical data.
These models are usually based on reasonable physical
assumptions about the nature of the particle source,
and/or the influence of the different objects placed in the
pathway of the particle beam in the experimental setup.
Here, we have shown how the two--slits experiment with cold neutrons
performed by Zeilinger {\it et al.} \cite{Zeilinger1} can be
explained to a certain extent by means of a simple analytical model
based on such kind of considerations.
This analysis has allowed us to characterize the relevance of the
different sources of incoherence present in the experiment.

However, these procedures mask the presence of another relevant
phenomenon that is likely to exist, and may influence importantly
the experiment: decoherence.
This mechanism, also leading to the suppression of coherence in the
system, cannot be described by optical models.
Decoherence is produced by the dynamical interaction between system
and environment (e.g., by means of scattering events), which is not
included in such models.

In this paper we have shown that decoherence is likely to
exist in Zeilinger {\it et al.}'s experiment, and that it can
explain fairly well with a simple model assuming an exponential
damping of the interference term \cite{Savage},
and using the experimental value of the coherence degree
($\Lambda_t = 0.63$).
By means of this model we have been able to establish the
influence of each element of the experimental arrangement on the
neutron beam.
Among these influences, the most important one is the effect that
the double--slit causes on the neutron beam: it splits the initial
quasi--plane wavefront coming from $C$ into two coherent Gaussian beams.


\begin{acknowledgments}
F.B. and M.J.B. gratefully acknowledge Prof.\ T.\ Seligman for his
hospitality at the Cuernavaca CIC (UNAM, Mexico), where this
collaboration was started.
A.S.S.\ is deeply indebted to Prof.\ Paul Brumer for interesting
discussions on the ideas developed here.
This work was supported in part by MCyT (Spain)
under contracts BFM2000--347 and BQU2003--8212.
A.S.S. gratefully acknowledges a doctoral grant from the Consejer\'\i a
de Educaci\'on y Cultura of the Comunidad Aut\'onoma de Madrid (Spain).
\end{acknowledgments}



\begin{thebibliography}
\bibitem{note1}
 By ``stationary'' we mean that the time average of these processes
 is independent on the choice of the time origin.
\bibitem{note3}
 We use the term ``correlation pattern'' because, as inferred from
 Eq.~(\ref{eq:7}), the modulation appears in the correlations between
 two different positions $x_1$ and $x_2$ on $O$, but not for $x_1 = x_2$.
 This is a consequence of considering $S$ as an incoherent source.
 Nonetheless, the pass through $O$ leads to assume certain coherence
 for the neutron beam [see Eq.~(\ref{eq:11}), for example].
\bibitem{note4} The inclusion in the calculations of some reasonable degree
 of non--coherence along the width of the slits do not change the results
 significantly.
\bibitem{note5}
 Time--dependence is indicated by a subscript ``$t$'' (e.g.,
 $|\Psi (t) \rangle = |\Psi\rangle_t$), while initial values
 do not carry any label (e.g., $|\Psi (0) \rangle = |\Psi\rangle$).
\end{thebibliography}
\end{document}